# The Discovery of Two Giant Arcs in the Rich Cluster A2219 with the Keck Telescope.*

Ian Smail,[1][†] David W. Hogg,[2] Roger Blandford,[2] Judith G. Cohen,[1] Alastair C. Edge[3] and S. George Djorgovski[1]

1) *Palomar Observatory, Caltech 105-24, Pasadena CA 91125*
2) *Theoretical Astrophysics, Caltech 130-33, Pasadena CA 91125*
3) *Institute of Astronomy, Madingley Rd, Cambridge CB3 0HA*

**ABSTRACT**
We report the discovery with the Keck telescope of two new multiply-imaged arcs in the luminous X-ray cluster A2219 ($z = 0.225$). The brighter arc in the field is red and we use spectroscopic and photometric information to identify it as a $z \sim 1$ moderately star-forming system. The brightness of this arc implies that it is formed from two merging images of the background source, and we identify possible candidates for the third image of this source. The second giant arc in this cluster is blue, and while fainter than the red arc it has a similarly large angular extent (32 arcsec). This arc comprises three images of a single nucleated source – the relative parities of the three images are discernible in our best resolution images. The presence of several bright multiply imaged arcs in a single cluster allows detailed modelling of the cluster mass distribution, especially when redshift information is available. We present a lensing model of the cluster which explains the properties of the various arcs, and we contrast this model with the optical and X-ray information available on the cluster. We uncover significant differences between the distributions of mass and X-ray gas in the cluster. We suggest that such discrepancies may indicate an on-going merger event in the cluster core, possibly associated with a group around the second brightest cluster member. The preponderance of similar merger signatures in a large fraction of the moderate redshift clusters would indicate their dynamical immaturity.

## 1 INTRODUCTION

Giant gravitational arcs are highly distorted images of distant field galaxies, formed by the strongly peaked mass concentrations in the centres of rich clusters. They can thus be employed to probe the distribution and amount of mass in rich clusters. The most useful lensed features for this sort of study are the multiply-imaged giant arcs and the new class of multiply-imaged pairs (Kneib et al. 1993, Smail et al. 1995a). The distinctive morphologies of these features make them easy to identify and the positions and relative magnifications of the various images can provide a fairly detailed view of the distribution of mass in the cluster core. In particular constraints can be derived on the shape and extent of the mass distribution in the cluster core.

In contrast to these strongly lensed features, individual weakly distorted arclets provide limited information about the cluster core (Kochanek 1990) and are significantly harder to recognise, especially in shallow exposures, resulting in more misidentifications (e.g. the "arcs" in Q0957+561 and MS1455+22). Nevertheless, if a sufficient number of these weakly distorted background images are available they can be employed in statistical tests (Kaiser & Squires 1993, Bonnet & Mellier 1994) to reconstruct a map of the projected mass distribution in the cluster on larger scales than that probed by the giant arcs (e.g. Bonnet et al. 1994, Fahlman et al. 1994, Smail et al. 1995b).

---

There is considerable interest in studying the mass distributions of rich clusters, primarily to understand the growth and evolution of structure in the universe. This interest has recently increased with the discovery of a surprisingly rapid decline in the number density of luminous X-ray clusters beyond $z \sim 0.2$ (Edge *et al.* 1990, Henry *et al.* 1992, Bower *et al.* 1994). This is widely assumed to result from hierarchical growth of clusters through mergers, although there are difficulties in recreating even the gross features of the observations using simple hierarchical cosmological models (e.g. Kaiser 1991). The bulk of the X-ray emission in these luminous clusters occurs within $\lesssim 500$ kpc of the cluster centre.[‡] Thus to understand the role of cluster growth and mergers in the X-ray evolution we need to study the mass distribution on similar scales. The giant arcs provide an unique probe of the central regions of rich clusters which can be compared with the results from X-ray imaging to give a clearer view of the evolution of rich clusters.

Comparisons of lensing and X-ray results have already provided some intriguing results on the relative distributions of light and mass in the very centres of rich clusters ($\lesssim 200$ kpc). Hammer & Rigaut (1989) pointed out that the properties of known giant arcs were difficult to reconcile with the mass distributions of rich clusters, as predicted from local X-ray studies. The X-ray observations indicated fairly large core radii in the mass ($r_c \sim 250$ kpc), while the narrow widths of the giant arcs were most easily explained if the mass was very centrally concentrated ($r_c \lesssim 50$–$100$ kpc). This argument relied upon the assumption that the background sources have scales sizes typical of local bright galaxies. Nevertheless, more general arguments using the high curvature of the majority of giant arcs give a similar conclusion (Miralda-Escudé 1993). Most of the detailed lensing models of rich clusters performed to date confirm the latter result and give values of the mass core radius of $r_c \lesssim 100$ kpc. Indeed, where a radial arc is present (MS2137$-$23, Mellier *et al.* 1993; A370, Smail *et al.* 1995c) the identification of the core radius is unambiguous. Tellingly, both these clusters have core radii of $r_c \sim 50$ kpc.

In this communication we present the discovery with the 10-m Keck telescope of two multiply-imaged giant arcs in the luminous X-ray cluster A2219. Section 2 details the results from our imaging and spectroscopic observations. In section 3 we discuss our modelling of the various lensed features in the cluster. Finally, in section 4 we contrast the results of our modelling with the distributions of the baryonic tracers in the cluster and give our main conclusions.

## 2  OBSERVATIONS

Abell 2219 is a luminous X-ray cluster at $z = 0.225$ (Allen *et al.* 1992). The cluster is part of a sample of the most luminous northern X-ray clusters in the redshift range $0.2 < z < 0.3$ selected from the ROSAT All Sky Survey. This sample is being used as the basis of an extensive X-ray and optical study of the evolution of rich clusters at moderate redshifts (Edge *et al.* 1995, Smail & Edge 1995).

### 2.1  Imaging Observations

The discovery image of the arcs in A2219 was acquired in mediocre observing conditions (seeing 1.2-1.5 arcsec) during the night of April 12-13 1994 using the Low Resolution Imaging Spectrograph in imaging mode (LRIS, Oke *et al.* 1995) on the 10-m Keck telescope on Mauna Kea, Hawaii. LRIS is designed as a double-armed (red & blue) imaging spectrograph with a 6×8 arcmin field of view, low to moderate spectral resolution and good spatial sampling (0.22 arcsec/pixel). Subsequently, a number of additional exposures in $UBVI$ of the cluster were taken with the COSMIC imaging spectrograph (Dressler *et al.* 1995) on the 5-m Hale telescope at Palomar during the nights of June 9-12 1994 and July 5-6 1994. Normal IRAF reduction procedures were used to process both the object frames and standards. The final stacked images comprise 3.7 ksec in $U$ with 1.6 arcsec seeing, 1.25 ksec in $B$ in 1.3 arcsec seeing, 1.0 ksec in $V$ in 1.0 arcsec seeing and 0.5 ksec in $I$ with 1.2 arcsec seeing. The $1\sigma$ surface brightness limits are: $\mu_U = 27.8$, $\mu_B = 28.4$, $\mu_V = 27.6$ and $\mu_I = 25.9$. The galactic extinction towards A2219 is $A_B = 0.0$ and so no reddening corrections have been applied.

#### 2.1.1 The Cluster Galaxies

We show in Figure 1 the $I$ band exposure of the cluster centre. The cluster is easily seen as an elongated structure, whose centre is dominated by a giant cD galaxy ($z = 0.225$, Allen *et al.* 1992) with a stellar envelope extending to radii of at least 150 kpc. The central region of the cluster is extremely rich, with 24 galaxies discernible within 100 kpc of the cD. The total integrated luminosity within this radius is $L_V \sim 1.1\ 10^{12} L_\odot$. A secondary concentration of galaxies is visible around the second brightest cluster member (AEF-2, $z = 0.2344$, Allen *et al.* 1992) at a projected radius of 230 kpc from the cD with a position angle of $\theta = 237$ degrees. Diffuse light apparently connects this sub-clump with the envelope of the central cluster galaxy (Figure 2), although there is a $c\Delta z \sim 2300$ km/sec velocity offset between the two components (Allen *et al.* 1992).

From our multi-colour photometry we can select galaxies with colours similar to those expected for a spheroidal system at $z \sim 0.23$ (Figure 3) to estimate the optical richness of the cluster. There are 49 galaxies within 500 kpc of the cD with magnitudes in the range $[m_3, m_3 + 2]$ giving A2219 an optical richness comparable to Coma (Metcalfe 1983). Fitting a modified Hubble Law to the radial distribution of cluster members we derive a core radius of $r_c \sim 220^{+120}_{-140}$ kpc. The distribution of colour-selected cluster members is centred on the cD with an elliptical distribution, position angle $\theta = 70 \pm 2$ degrees, close to the orientation of the major axis of the cD's envelope.

#### 2.1.2 The Candidate Arcs

Two arc-like structures are clearly visible around the central cluster galaxy in Figure 2. The longer, fainter arc ($L$) lies to the west of the cD, while the brighter, short arc ($N_{1+2}$) lies to the east. $L$ comprises three separate segments ($L_1$, $L_2$ & $L_3$) identified with a peak in the arc surface brightness and with a break between each (Figure 2,

---

[‡] We take $H_0 = 50$ kms/sec/Mpc and $q_0 = 0.5$ unless otherwise stated. In this cosmology 1 arcsec corresponds to 4.58 kpc at $z = 0.225$.

Table 2). All three segments are either unresolved or very marginally resolved on our best seeing $V$ images (Table 1). The candidate arc $N_{1+2}$ also appears to consist of two sections, although this is not so obvious. The intrinsic width of arc $N_{1+2}$ shows it to be nearly resolved in our best seeing images (Table 1).

To measure magnitudes for the arcs all frames were aligned, convolved to the seeing of the worst image and then magnitudes measured within elliptical apertures whose major and minor diameters are listed in Table 1. The background contribution was estimated by median-interpolating across these apertures. The two arcs $L$ and $N_{1+2}$ have very different combinations of broad-band colours (Table 1). Both arcs have blue uv-optical colours and contrast strongly against the cluster ellipticals in the $U$ exposure of the cluster. Their $(U-B)$ colours are in the bluest ~1% of objects in the field at their apparent magnitudes. In contrast to this, their optical colours are very different. In particular, in $(V-I)$ arc $L$ is very blue, while arc $N_{1+2}$ is even redder than the cluster ellipticals (Figure 3). Figure 3 also shows the observed $(B-V)$ and $(V-I)$ colours for all galaxies in the field brighter than $I = 23$. In addition we plot the loci of various local galaxy spectral energy distributions (SED) as a function of redshift to indicate the variation in observed colour from both star-formation rate and K-corrections.

All three sections of arc $L$ share the same $(U-B)$, $(B-V)$ and $(V-I)$ colours supporting our assertion that they are all images of a single background source. Combining the three components we see that arc $L$ has colours close to those of a strongly star-forming galaxy (Figure 3), typical of the highest redshift arcs (Smail *et al.* 1993). In contrast, the red $(V-I)$ colour of arc $N_{1+2}$ probably indicates that a break exists in the galaxy's SED between the $V$ and $I$ bands. Indeed, if we look at the joint $(B-V)$–$(V-I)$ colour of $N_{1+2}$ we see it is most consistent with a moderate star-formation SED (Scd-type) at $z \sim 1$ (Figure 3), although a bluer SED at $z \gtrsim 0.6$ would also fit. There is no strong evidence for colour variations within the various components of either arc.

The brightness of arc $N_{1+2}$ indicates that it might comprise two merging images of a single background source. A detailed inspection of our $B$ and $V$ images of the arc also hints at bimodal structure. If $N_{1+2}$ does consist of two images then we would expect a third, fainter image of the source to the east or north-east of the central galaxy. A number of faint objects are seen around the bright galaxy $C$ in Figure 2. We used the blue $(U-B)$ and $(B-V)$ colours of $N_{1+2}$, in conjunction with its red $(V-I)$ colour, to select possible third images. The best candidate is marked as $N_3$ in Figure 3 and colours are given in Table 1. The overall shape of the object's SED is consistent with that determined for $N_{1+2}$ and moreover the peak surface brightness of the source is similar to that seen in $N_{1+2}$. We feel the evidence is convincing enough to state that $N_{1+2}$ is formed from two merging images of a background source, with $N_3$ as the third image.

## 2.2 Spectroscopic Observations

To verify that some of the features we observed in A2219 are indeed gravitationally lensed we have obtained a spectrum of the candidate arc $N_{1+2}$ with the LRIS spectrograph on Keck. These spectra were taken on the night of August 9-10 1994 and comprise a total of 6.0 ksec integration on $N_{1+2}$ using a long slit aligned along the major axis of arc. The slit width was 1.0 arcsec, with the 300 line/mm grating this gave a resolution of 2.46Å/pixel across the wavelength range $\lambda \sim 3950$—9000Å. The data was reduced, extracted and fluxed using standard FIGARO procedures. The long slit passed through a number of serendipitous objects in addition to the $N_{1+2}$, including the halo of the bright galaxy $C$ in Figure 2. This galaxy shows [OIII]5007 and has a redshift of $z = 0.2393$, $c\Delta z \sim 3500$ km/sec relative to the cD.

We show the final sky-subtracted spectrum of arc $N_{1+2}$ in Figure 4. We obtained reasonable signal to noise (s/n~ 5) in the continuum of the arc given the relatively short exposure time. However, we do not detect any strong emission features in the spectral region $\lambda \sim$3950–9000Å. At the redder end of the spectral range this conclusion is based upon visual inspection of the raw and sky-subtracted two dimensional spectra. The absence of emission lines blueward of ~6100Å, coupled with the blue uv-colours of the arc effectively rules out identification of the object as either a cluster member or foreground galaxy. However, a break in the continuum is visible longward of $\lambda \sim 7500$Å. To confirm this we median filtered the spectrum using an algorithm which rejects regions where there is strong sky emission. This smoothed spectrum is shown in Figure 4 and confirms the presence of a sharp upturn in the region $\lambda \sim 7600$–8000Å. Identification of this feature with the 4000Å break yields a redshift of $z \sim 0.9$–1 for the source. At this redshift the strongest emission line we could hope to detect would be [OII]3727, which would lie around $\lambda \sim 7500$Å, spanning the atmospheric A-band. We see no emission line in the arc around this wavelength and determine a conservative limit on the observed equivalent width of the line of EW([OII])$\lesssim$ 24Å. In the local population this limit on the [OII] emission would give the galaxy an SED similar to a mid-type spiral (Kennicutt 1992).

Both the spectroscopic redshift of the arc and the apparently modest star-formation are consistent with the redshift range and identification based upon the broad-band colours. We therefore conclude that the source seen as arc $N_{1+2}$ is: (1) definitely beyond the cluster and (2) most likely a galaxy at $z \sim 1$ with modest star-formation.

## 2.3 X-ray Observations

Abell 2219 has an extremely high X-ray luminosity of $L_X(0.1 - 2.4) = 1.8 \ 10^{45}$ ergs sec$^{-1}$. A 13.4 ksec ROSAT HRI exposure of the cluster from Edge *et al.* (1995) is shown in Figure 1. The cluster emission is well detected out to ~1 Mpc with a total of ~2700 counts. A serendipitous point source 5.3 arcmin east of the cluster lies within 4 arcsec of a $13^m$ star, defining the pointing accuracy of the observation. The X-ray morphology in the central regions is highly elliptical (Table 4), centred close to the central cluster galaxy. There is no observable twist to the X-ray isophotes across the range quoted in Table 4 and no indication of emission associated with the second brightest cluster member. There is, however, marginal evidence for a number of faint point sources ($L_X(0.1 - 2.4) = 10^{42}$–$10^{43}$ ergs sec$^{-1}$) in the clus-ter core which may be associated with the starburst/UVX galaxies seen there.

By assuming an isothermal temperature distribution for the cluster gas we can deproject the X-ray image to determine the cluster mass profile, more details of this procedure are given in Edge *et al.* (1995). The best fit core radius is $r_c \sim 310^{+120}_{-80}$ kpc. The relatively shallow central X-ray surface brightness profile suggests that there is no strong cooling flow in this cluster. Our deprojection analysis confirms this statement, we derive a cooling flow of $\dot{M} = 100$–$300$ $M_\odot/yr$. This flow rate is smaller than those determined for other highly luminous X-ray clusters, which predict $\dot{M} = 500$–$1500$ $M_\odot/yr$. The central cluster galaxy also shows no strong line emission supporting the absence of a strong cooling flow (Allen *et al.* 1992).

## 3 MODELLING AND RESULTS

We have identified a number of gravitationally lensed features in the central regions of the rich cluster A2219. We now propose to model these features to study the mass distribution in the cluster and to compare this to the distribution inferred from indirect, baryonic tracers of the cluster mass.

Our gravitational lens models consist of a smooth elliptical "cluster" mass distribution centred on the cD galaxy plus concentrated mass on those bright galaxies close to the arc images which may influence their locations. The cluster is modeled with a scaled 2-dimensional potential $\psi(x, y)$, where $x$ and $y$ are angular coordinates increasing to the east and north. The potential $\psi$ is related to the mass by $\nabla^2 \psi = 2\kappa$, where $\nabla^2$ is the two-dimensional Laplacian operator on the lens $(x, y)$ plane and $\kappa$ is the ratio of the surface density of the lens to the "critical surface density" for lensing (e.g. Blandford & Narayan, 1992 or Schneider, Ehlers & Falco, 1992). The high ellipticity shown by the baryonic tracers in the cluster forces us to adopt a novel form for the lensing potential which can adequately describe such morphologies. We construct the cluster potential by superimposing potentials of the form:

$$\psi(x,y) = \frac{b^{2-2q}}{2q} \left\{ s^2 + (1+\epsilon_c)\, x^2 + 2\,\epsilon_s\, x\, y + (1-\epsilon_c)\, y^2 \right\}^q \quad (1)$$

where $b$ is the angular critical radius, $s$ is the angular core radius, $\epsilon_c$ and $\epsilon_s$ are related to the ellipticity $\epsilon$ and position angle $\theta$ of the composite potential by $\epsilon_c = \epsilon \cos 2\theta$ and $\epsilon_s = \epsilon \sin 2\theta$, and $q$ is an exponent $\leq 1/2$. The use of single potentials of the form (eq. 1) leads to peanut-shaped surface density contours when $\epsilon > 0.2$ and $q = 1/2$, and this problem is more acute for smaller $q$ or larger $\epsilon$ (Blandford & Kochanek, 1987; Kassiola & Kovner, 1993). Superposition of two or more suitably chosen potentials with this form obviates this problem at little numerical cost. We find that a superposition of an elliptical potential with $q < 1/2$ and a spherical one with $q = 1/2$ is adequate for our purposes here. The final composite potential is not strictly isothermal in the cluster centre, although it converges to an isothermal form at large radii. The additional cluster galaxies are modeled with point masses. We constrained all these galaxies to have roughly the same mass-to-light ratio, except for galaxy $A$ (see below).

To find a model that fits the observations, we choose an initial set of parameters for the lens model, subject to the constraints available from arc $N$, and a position, length and orientation for the source of $L$ on the source plane. We then use a downhill-simplex method (Press *et al.* 1992) to minimize $\chi^2$ between the observed and predicted features of the images of $L$ on the image plane. We define

$$\chi^2 = \sum_i \frac{(O_i - M_i)^2}{\sigma_i^2}, \quad (2)$$

where the $O_i$ are the observed quantities, the $\sigma_i$ are the associated observational errors, and the $M_i$ are the model values. We use a Newton-Raphson method to locate the images of $L$ and their lengths and orientations. In addition we determine the ratio

$$\beta = \frac{D_N\, D_{dL}}{D_L\, D_{dN}}, \quad (3)$$

where $D_L$ and $D_N$ are angular diameter distances from observer to the sources of $L$ and $N$, and $D_{dL}$ and $D_{dN}$ are angular diameter distances from lens (deflector, $d$) to the sources, such that $N$ is on a critical curve. This also provides us with a source location for $N$ and we use Newton-Raphson method again to locate the third image $N_3$. With liberal positional errors of $0.5''$, we find a range of models that fit the positions, orientations and lengths of the three images of $L$ and roughly reproduce the position of the extra image $N_3$.

Parameters of a typical model (the "fiducial model") are given in Table 3 and the potential contours, surface density contours and critical curve for $L$ are shown in Figure 5. The uncertainties quoted in Table 3 are not formal errors but subjective estimates of the parameter ranges among qualitatively similar successful models. Parameters with no uncertainties given were not varied in our modelling.

The fiducial gravitational lens model puts the sources of $L$ and $N$ inside "naked cusps" (e.g. Schneider *et al.* 1992), so there are three images of each source (counting $N_{1+2}$ as a double) with no counter-arcs visible. This is not an important feature of the models, as five-image models can be found if the core radius is decreased. Since the core radius is not strongly constrained, we do not rule out the eventual discovery of counter-arcs with deep, high-resolution imaging. Such a discovery would put strong constraints on the core radius of the mass distribution, we discuss later another approach which would constrain the core radius of the mass distribution in the cluster core. The most significant shortcoming of the fiducial model is the predicted position of $N_3$ which lies $\sim 4$ arcsec from the observed position. We do not view this as a failure of the model but rather attribute it to small scale structure in the cluster. A number of avenues are open to us to solve this problem including more realistic individual galaxy potentials (with finite extent and individual ellipticities) or a relaxation of the constraint that all galaxies have the same mass-to-light ratio, but these options open up model space dramatically. The mass to light ratios determined for the cluster galaxies in the fiducial model are effective values within the distance between a particular cluster galaxy and the nearest segment on an arc. The median aperture is then $\sim 90$ kpc in diameter within which we require a mass to light of $M/L \sim 40 (M_\odot/L_\odot)_V$. We need a higher mass to light around galaxy $A$ than the remainder of the cluster galaxies used, in order to fit the position and ori-entation of $L_3$. The high derived mass of galaxy $A$ may in fact be just compensating for the simplicity of our cluster potential model, which does not allow for octopole or higher-order distortions, or any breaking of mirror symmetry. Using a more complex distribution for the cluster potential could significantly reduce the mass in the vicinity of galaxy $A$, as well as allowing us to fit $N_3$.

Any model of the cluster mass will only be well constrained within the radii of the giant arcs. We have therefore chosen to only quote values for the integrated mass within this radius, where our results are robust to gross changes to the form of the cluster profile. Our mass model then gives a mass in a 100 kpc radius cylinder through the cluster centre of $M \sim 1.6 \times 10^{14} M_\odot$, combined with the integrated luminosity in the same aperture we derive a mass to light ratio of $M/L \sim 145 (M_\odot/L_\odot)_V$, comparable to that measured using giant arcs in other clusters (e.g. Kneib *et al.* 1995a). Moreover, this $M/L$ is similar to that required in our model in the vicinity of galaxy $A$ and supports our contention that mass distribution in the cluster core has complex structure on scales of $\sim 100$ kpc.

We can translate our mass estimate into a crude velocity dispersion, $\sigma$, for the cluster assuming the mass distribution is roughly isothermal. We use the redshift of arc $N$ and its projected radius in the cluster, $b_N$ in

$$\sigma^2 \approx \frac{c^2}{4\pi} \frac{D_s}{D_{ds}} b_N, \qquad (4)$$

where $D_s$ and $D_{ds}$ are angular diameter distances from observer to source and lens to source (in this case $N$). Our fiducial model then predicts $\sigma \sim 930$ km/sec. As the cluster is highly elliptical and arc $N$ straddles the major axis of the mass distribution this dispersion is likely to be an upper limit to the true value. To date there are only redshifts of a few cluster members (*c.f.* Allen *et al.* 1995) and so we are unable to determine a galaxy velocity dispersion directly. Alternatively, if we assume a roughly isothermal gas distribution in equilibrium with the cluster potential we can translate our mass estimate into a prediction for the temperature of the cluster gas giving $T \sim 7.9$ KeV. ASCA observations of this cluster have been recently obtained and the comparison of our prediction with the observed gas temperature will directly address the current controversy about mass estimates of cluster cores from X-ray observations (*e.g.* Babul & Miralda-Escude 1994).

In our fiducial model the source of $N$ is not at the same redshift as that of $L$. For each model we have an associated value of $\beta$, a ratio fixed by the arc redshifts and world model, so each model makes a prediction for $z_L$. The fiducial model predicts $z_L \sim 2.0$ but we find that the prediction is model-dependent and almost any redshift $z_N < z_L$ is allowed. As the predicted redshift of $L$ depends on the world model, in principle an exhaustive search of model space plus a measurement of $z_L$ could put limits on the cosmological parameters $\Omega$ and $\Lambda$. However, in practice the uncertainty in the lens modelling is likely to be much greater than the variation due to cosmology.

## 4 DISCUSSION AND CONCLUSIONS

We list in Table 4 the basic morphological information on the cluster mass from our preferred lens model. As we have stated this model does not provide a detailed description of all the lensed features within the cluster core. We feel, however, that it is sufficiently close that the gross properties of the model must be correct. Table 4 also lists the basic morphological information of the baryonic tracers in the cluster centre: the stellar halo of the cD and the hot X-ray gas. In the case of the X-ray gas we quote the ellipticity of the surface brightness distribution, for an isothermal gas distribution the ellipticity of the underlying mass distribution is roughly three times this. Thus if we make the usual assumptions that the X-ray gas is both in hydrostatic equilibrium and isothermal it would indicate an extremely elliptical mass distribution.

Comparing just the mass model and the halo of the cD we see very good agreement between the orientations and ellipticities of these two distributions. Similar conclusions have previously been reached for other clusters (*e.g.* Kneib *et al.* 1995a), but in the majority of these cases it was imposed onto the models as a rigid constraint. The agreement shown in A2219 is therefore more striking given the complete freedom allowed in the mass modelling. The similarities of the stellar envelope of the central galaxy to the mass distribution indicate that the envelope is fully relaxed in the dark matter potential. If this is the case then the stellar velocity dispersion in the halo must rise from its value in the centre of the cD, probably $\sigma \sim 300$–$400$ km/sec, to the value determined at the radius of the arcs, $\sigma \sim 950$ km/sec. Miralda-Escudé (1994) has demonstrated how the dispersion profile of the stellar envelope can be used to place a lower limit on the core radius of the dominant dark matter component in the cluster centre. It is thus possible to observationally confirm the small core radius indicated by our lens model.

If we extrapolate the shape of the mass distribution in the cluster centre out to slightly larger radii we can compare it with the morphology of the X-ray map on similar scales. This comparison results in a striking conclusion (Table 4), the major axis of the X-ray gas is misaligned with that inferred for the mass distribution (and with orientation of the number-weighted galaxy distribution on similar scales). The deviation $\Delta\theta \sim 13$ degrees is significantly larger than the estimated error, $\sim 4$ degrees. Taken in conjunction with the extreme ellipticity of the X-ray gas emission and the large inferred core radius, this indicates that the cluster gas can not be in hydrostatic equilibrium in the cluster potential. Searching for a cause of the misalignment we note that the orientation of the major axis of the X-ray emission lies close to the direction between the cD and the compact group of galaxies around the second brightest cluster member. We suggest that the cluster is suffering, or has recently undergone, a core-penetrating merger by a group of galaxies associated with the second brightest cluster member. This provides a natural explanation of the misalignment of the X-ray gas with the mass distribution as the merging group injects kinetic energy into the cluster gas along the merger axis. The gas will relax back into equilibrium in the cluster potential on the time scale of a few crossing times ($\sim 1$–$2$ Gyr). The presence of diffuse light connecting the cD with this group supports the conjecture that the sub-clump is physically close to the cluster core. A very simple dynamical model using the observed velocity offset between the central galaxy and the second brightest cluster member, $c\Delta z \sim 2300$ km/sec, and our estimate for the central mass of the cluster shows that the sub-clump is likely to be in a bound orbit and will eventually merge with the cluster core.

The importance of the misalignment of the gas and mass distributions is that it is a clear signature of that a merger event has occurred within the last $\sim 1$–$2$ Gyr (Evrard 1990). Our ability to assign a timescale to the merger makes this signature a more powerful probe of cluster growth than classical substructure tests (A.I. Zabludoff priv. comm.). When using the rate of occurence of spatial or velocity substructure within the cluster galaxy distribution the lifetime of any feature depends critically on structural parameters of the cluster and sub-clump, as well as the orbital characteristics of the sub-clump. These additional factors complicate comparison of observations with theoretical models when trying to constrain in-fall rates into clusters and thence cosmological parameters (Richstone *et al.* 1992). A detailed discussion of the role of such mergers in the evolution of distant, luminous X-ray clusters requires lensing observations of a better defined cluster sample than currently exists. Nevertheless, it is already apparent that merger signatures, seen as discrepancies between the mass and X-ray distributions, are relatively common features in rich, moderate redshift clusters (*e.g.* Kneib *et al.* 1995, Smail *et al.* 1995b). Searches for similar merger signatures in local clusters are critical for understanding in-fall into clusters and these are already underway (Zabludoff & Zaritsky 1995).

Future observations of A2219 with large telescopes should provide us with a more secure redshift for $N$ and also a redshift for $L$. With very high-resolution imaging using HST we can better locate and characterize the multiple images, verify our association of $N_3$ with $N_{1+2}$, and constrain model space enormously. Such imaging would also allow a search for counter-arcs and fainter multiple-image systems. Furthermore, HST observations of distorted background galaxies ("arclets") would put strong constraints on the detailed form of the mass distribution in the cluster core (Kneib *et al.* 1995b). Future observations of A2219 and other similarly well-constrained lenses will quantify the role of mergers in the growth and evolution of rich clusters.

## ACKNOWLEDGEMENTS

We thank Bev Oke for leading the construction of the LRIS spectrograph. Jim McCarthy and Jim Westphal are thanked for their work in providing an efficient $U$ imaging capability on the Hale 5-m and Jim Schombert for providing his $U$ filter. We acknowledge helpful discussions with Richard Ellis, Jean-Paul Kneib, Jordi Miralda-Escudé and especially Ann Zabludoff. Support via a NATO Advanced Fellowship (IRS), NSF Graduate Fellowship (DWH), NSF grant AST92-23370 (RDB) and partial support from a NSF PYI award AST-9157412 (SGD). is gratefully acknowledged. Finally, it is a pleasure to thank the W.M. Keck Foundation and its President, Howard B. Keck, for the generous grant that made the W.M. Keck Observatory possible.

## REFERENCES

Allen, S.W., Edge, A.C., Fabian, A.C., Böhringer, H., Crawford, C.S., Ebeling, H., Johnstone, R.M., Naylor, T. & Schwarz, R.A., 1992, MNRAS, 259, 67.
Allen, S.W., *et al.*, 1995, in prep.
Babul, A. & Miralda-Escudé, J., 1994, MNRAS, submitted.
Blandford, R.D. & Kochanek, C.S., 1987, ApJ, 321, 658.
Blandford, R.D. & Narayan, R., 1992, ARAA, 30, 311.
Bonnet, H., Mellier, Y. & Fort, B., 1994, ApJL, 427, L83.
Bonnet, H. & Mellier, Y., 1994, A&A, submitted.
Bower, R.G., Bohringer, H., Briel, U.G., Ellis, R.S., Castander, F.J. & Couch W.J., 1994, MNRAS, 268, 345.
Dressler, *et al.*, 1995, in prep.
Edge, A.C., Stewart, G.C., Fabian, A.C. & Arnaud, K.A., 1990, MNRAS, 245, 559.
Edge, A.C., Allen, S.W., Fabian, A., White, D.A. & Böhringer, H., 1995, in prep.
Evrard, A.E., 1990, ApJ, 363, 349.
Fahlman, G., Kaiser, N., Squires, G. & Woods, D., 1994, ApJ, 437, 56.
Hammer, F. & Rigaut, F., 1989, A&A, 226, 45.
Henry, J.P., Gioia, I.M., Maccacaro, T., Morris, S.L., Stocke, J.T. & Wolte, A., 1992, ApJ, 386, 408.
Kaiser, N., 1991. ApJ, 383, 104.
Kaiser, N. & Squires, G., 1993, ApJ, 404, 441,
Kassiola, A. & Kovner, I., 1993, ApJ, 417, 450.
Kennicutt, R.C., 1992, ApJ, 388, 310.
Kneib, J-P., Mellier, Y., Fort, B. & Mathez, G., 1993, A&A, 273, 367.
Kneib, J-P., Mellier, Y., Pelló, R., Miralda-Escudé, J., Le Borgne, J-F., Böhringer, H. & Picat, J-P., 1995a, A&A, submitted.
Kneib, J-P., Ellis, R.S., Smail, I., Couch, W.J. & Sharples, R.M., 1995b, Nature, submitted.
Kochanek, C.S., 1990, MNRAS, 247, 135.
Mellier, Y., Fort, B. & Kneib, J-P., 1993, ApJ, 407, 33.
Miralda-Escudé, J., 1993, ApJ, 403, 497.
Miralda-Escudé, J., 1994, IAS preprint.
Metcalfe, N., 1983, PhD thesis, Univ. of Oxford.
Oke, J.B., Cohen, J.G., Carr, M., Cromer, J., Dingizian, A., Harris, F.H., Labreque, S., Lucinio, R., Schaal, W., Epps, H. & Miller, J., 1995, PASP, in press.
Press, W.H., Teukolsky, S.A., Vetterling, W.T. & Flannery, B.P., 1992, Numerical Recipes in C, 2 ed., Cambridge University Press, Cambridge.
Richstone, D., Loeb, A. & Turner, E.L., 1992, ApJ, 393, 477.
Schneider, P., Ehlers, J. & Falco, E.E., 1992, Gravitational Lenses, Springer-Verlag, Berlin.
Smail, I., Ellis, R.S., Aragòn-Salamanca, A., Soucail, G., Mellier, Y. & Giraud, E., 1993, MNRAS, 263, 628.
Smail, I., Couch, W.J., Ellis, R.S. & Sharples, R.M., 1995a, ApJ, in press.
Smail, I., Ellis, R.S., Fitchett, M.J. & Edge, A.C., 1995b, MNRAS, in press.
Smail, I., Dressler, A., Kneib, J.-P., Ellis, R.S., Couch, W.J., Sharples, R.M. & Oemler, A., 1995c, in prep.
Smail, I. & Edge, A.C., 1995, in prep.
Zabludoff, A.I. & Zaritsky, D., 1995, preprint.

## Figures

**Figure 1:** An $I$ image of A2219 showing the distribution of bright galaxies around the central cD ($z = 0.225$). Overlayed upon this is the ROSAT HRI exposure, smoothed with a 70 kpc FWHM gaussian. This is peaked close to the cluster cD and shows a highly elliptical morphology which is misaligned with the stellar halo of the cD on smaller scales (Figure 2).

A number of 3–4$\sigma$ point source features are visible in the cluster centre, these may be associated with starburst/UVX galaxies within the cluster core. North is top, East is left.

**Figure 2:** The combined Hale and Keck $V$ exposure of the central regions of the cluster. The various components of the candidate arcs $L$ and $N$ are marked. The orientation is the same as Figure 1. At the cluster redshift 1 arcsec corresponds to 4.58 kpc.

**Figure 3:** ($B$–$V$)–($V$–$I$) plot of all objects in an 8×8 arcmin field centred on A2219 brighter than $I = 23$. We also plot the predicted colours of various local galaxy SED's as a function of redshift from $z = 0.0$ (left end of track) to $z = 1.2$ (right end of track), the + show $dz = 0.1$ increments. The two bold points mark the integrated colours and associated errors of $L$ and $N_{1+2}$. Notice, that while $L$ has the blue colours typical of the highest redshift arcs, $N_{1+2}$ is fairly red in ($V-I$) and can be quite well described by a galaxy with an SED similar to a present-day Scd spiral at $z \sim 1$.

**Figure 4:** The LRIS spectrum of the red arc $N_{1+2}$ in A2219. The bottom spectrum shows the sky-subtracted and fluxed spectrum, while the upper spectrum shows the continuum shape produced by median-filtering the spectrum with rejection of pixels dominated by strong night-sky emission. The discontinuity in the spectral shape in the region $\lambda \sim 7700$–8100Å is easily seen and we identify this as the 4000Å break giving a source redshift of $z \sim 1$.

**Figure 5:** The potential contours (solid, light), critical curve (solid, dark), and surface-density contours (dotted, light) for the fiducial model (see text and Table 3). The open squares mark the observed positions of the peaks of the images of $L$, the open circle marks the observed position of $N_3$, the open triangles mark the centers of individual galaxies included in the model, and the derived images appear as scattered dots (from a simple Monte-Carlo ray-tracing routine that assumes circular sources). Notice that the predicted position of $N_3$ is a few arcseconds west of the observed position. All galaxies except $A$ at $[16.7, -6.9]$ have the same mass-to-light ratio. There are several extra images of $N$ but these are artifacts of the point-mass model for the galaxies.

## Tables

**Table 1:** Photometry of the candidate arcs in A2219. These are aperture values calculated from seeing-matched images. The aperture sizes are given in the second column of the first table. The second table includes deconvolved estimates for the arc widths in addition to the peak surface brightnesses of the various arc components. Also listed are the colours measured for the spheroidal cluster population.

**Table 2:** Positions of lensed features. The positions of the various features are measured relative to the cD galaxy on the $V$ image (Figure 2). The position of the cD is $\alpha(2000) = 16^h 40^m 20.6^s$ $\delta(2000) = 46°42'40''$.

**Table 3:** Parameters and predictions of the fiducial gravitational lens model of A2219. The cluster is constructed from the superposition of a spherical and elliptical part. The positions of the individual galaxies are shown in Figures 2 and 5.

**Table 4:** Comparison of cluster parameters from the baryonic tracers and from the lensing models. The "scale" gives the spatial range over which the measurements apply.